\documentclass[aps,prd,reprint,superscriptaddress,showpacs,nofootinbib,longbibliography]{revtex4-1}
\usepackage{graphicx}
\usepackage{amsfonts}
\usepackage{amsmath}
\usepackage{mathrsfs}
\usepackage{epsfig}
\usepackage{slashed}
\usepackage{dcolumn}%

\usepackage{ulem}
\usepackage{color}
\usepackage{caption}
\usepackage{subcaption}
\usepackage{hyperref}

\def\sige{\sigma_{\chi,e}}

\def\nuflub8{\phi^\nu_B}
\def\nuflube7{\phi^\nu_{Be}}

\def\mdm{m_{\chi}}
\def\mev{m_\mathrm{evap}}

\def\lesim{\lesssim}
\def\grsim{\gtrsim}

\def\vesc{v_{\mathrm{esc}}}

\def\vdm{v_{\chi}}

\def\dPobs{\dot{P}_{\mathrm{obs}}}

\def\dPscm{\dot{P}_{\mathrm{SCM}}}

\def\apj{Astrophys.\ J.\ }

\def\physrep{Phys.\ Rept.\ }
\def\prd{Phys.\ Rev.\ D\ }

\def\jcap{J.\ Cosmol.\ Astropart.\ Phys.\ }

\def\TeV{\,\mathrm{TeV}}
\def\GeV{\,\mathrm{GeV}}
\def\MeV{\,\mathrm{MeV}}

\def\MV\TeV{\,\mathrm{MV}}
\def\cm{\,\mathrm{cm}}

\newcolumntype{p}{D{,}{\pm}{-1}}

\begin{document}



\title{Comment on ``Probing the Dark Matter-Electron Interactions via Hydrogen-Atmosphere Pulsating White Dwarfs''}

\author{Jia-Shu Niu}
\email{jsniu@sxu.edu.cn}
\affiliation{Institute of Theoretical Physics, Shanxi University, Taiyuan, 030006, China}

\author{Hui-Fang Xue}
\affiliation{Department of Physics, Taiyuan Normal University, Taiyuan, 030619, China}

\date{\today}

\begin{abstract}
  In \href{https://journals.aps.org/prd/abstract/10.1103/PhysRevD.98.103023}{Phys. Rev. D 98, 103023 (2018)}, a novel scenario was proposed to probe the interactions between dark matter (DM) particles and electrons, via hydrogen-atmosphere pulsating white dwarfs (DAVs) in globular clusters. The estimation showed that the scenario could hopefully test the parameter space: $5 \GeV \lesim \mdm \lesim 10^{4} \GeV$ and $\sige \grsim 10^{-40} \cm^{2}$, where $\mdm$ is the DM particle's mass and $\sige$ is the elastic scattering cross section between DM and electron. In this comment, we have determined the exact lower limit of the testable DM particle mass $\sim 1.38 - 1.58 \GeV$, which depends on $\sige$. This gives us a credible lower limit of the testable DM particle mass in above scenario, and provide a clear upper limit of the DM particle mass which we should consider in future research.
\end{abstract}

\maketitle

In our previous work \citep{Niu2018_dav}, we show that hydrogen-atmosphere pulsating white dwarfs (DAVs) in globular clusters could be used as the probes to detect the interactions between dark matter (DM) and electrons. The potential sensitivity on DM particle's mass ($\mdm$) and elastic scattering cross section between DM and electron ($\sige$) could be hopefully extended to a region $5 \GeV \lesim \mdm \lesim 10^{4} \GeV$ and $\sige \grsim 10^{-40} \cm^{2}$. The testable lowest mass (or the ``evaporation mass'' $\mev$) there ($5 \GeV$) was a conservative estimation based on the case that the evaporation effect could be ignored in the condition of Sun and DM-nucleon interactions \citep{Gould1987b,Gould1990a}. Qualitatively speaking, $\mev$ should be smaller in the case of DAV if we considered its stronger gravity potential, which makes it difficult for DM particles to evaporate. On the other hand, it should be larger if we considered the stronger interactions between electron than DM and nucleus, which makes it easier for DM particles to gain thermal kinetic energy from electrons. This should be calculated in details.

\begin{figure*}
\centering
\includegraphics[width=0.7\textwidth]{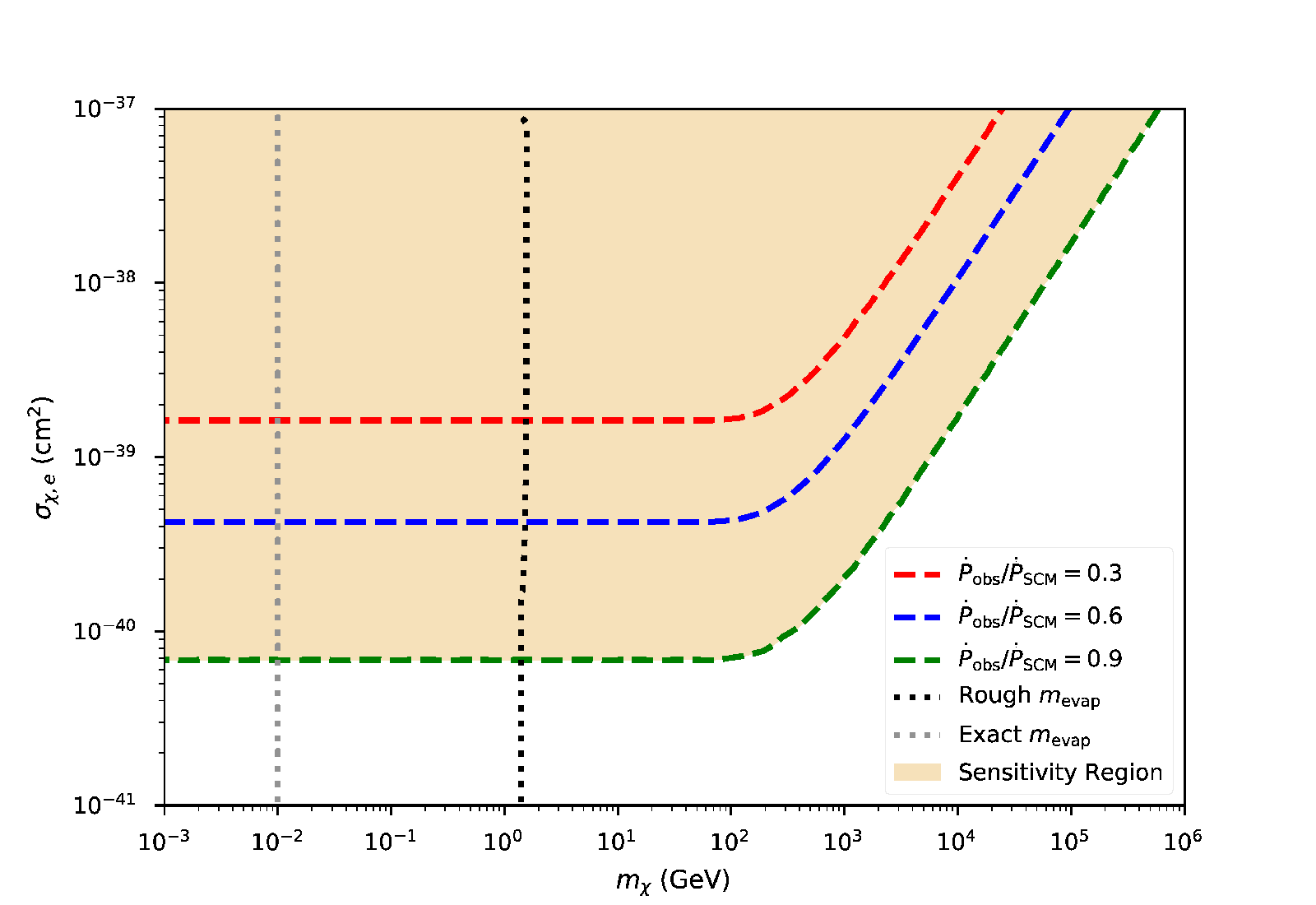}
\caption{The prospective constraints on $\sige - \mdm$ from the scenario in \citep{Niu2018_dav}. 
The lower limit of the exclusive line is determined by the uncertainty from $\dPobs/\dPscm$. The colored dashed lines correspond to specific $\dPobs/\dPscm$ values, and the grey and black vertical dotted lines represent the $\mev$ obtained in this comment, larger than which the evaporation effect could be ignored.}
\label{fig:sigma_mass}
\end{figure*}

Roughly speaking, we can estimate the evaporation mass by demanding that the typical velocity of a DM particle $\vdm = \sqrt{2 k T/ \mdm}$ (where $k$ is the Boltzmann constant and $T$ is the typical temperature in the star) be equal to the local escape speed of the star $\vesc$ \citep{Jungman1996,Hurst2015}. In this case, it gives the evaporation mass $\mev \sim 10 \MeV$, which is obviously smaller than the case in \citep{Niu2018_dav}\footnote{See in Fig. \ref{fig:sigma_mass}, the grey vertical dotted line.}. But it is obvious that in this case, most parts of the DM particles in the star are disappeared by evaporation rather than annihilation. As a result, the real value of $\mev$, which represents the lower limit of $\mdm$ when the evaporation effect should be ignored, should be larger.

Generally speaking, the evaporation mass should depend on the cross section between DM and the constituent of the star and the temperature distribution in it. Based on the work \citep{Garani2017}, following the detailed analytic expressions of the evaporation effect and the criterion to define $\mev$ in it\footnote{See Eq. (6.4) in \citet{Garani2017}}, we determine the evaporation mass when $10^{-41} \cm^{2} \lesim \sige \lesim 10^{-37} \cm^{2}$ and plot it in Fig. \ref{fig:sigma_mass} (the black vertical dotted line).

It shows that $\mev \sim 1.38 - 1.58 \GeV$, which is smaller than the estimation in \citep{Niu2018_dav} but much larger than the roughly estimation above. This gives us a credible lower limit of the testable DM particle mass in the  scenario proposed by \citep{Niu2018_dav}, and provide a clear upper limit of the DM particle mass which we should consider in future research.

\section*{Acknowledgments}
Jia-Shu Niu would like to appreciate Yi-Hang Nie and Jiu-Qin Liang for their trust and support.
This research was supported by the Special Funds for Theoretical Physics in National Natural Science Foundation of China (NSFC) (No. 11947125) and the Applied Basic Research Programs of Natural Science Foundation of Shanxi Province (No. 201901D111043).


%

\end{document}